\documentclass[prl,groupedaddress,superscriptaddress,reprint,twocolumn,showpacs,notitlepage]{revtex4}

\usepackage{amsfonts}
\usepackage{amsmath}
\usepackage{amssymb}
\usepackage{graphicx}
\usepackage{hyperref}
\usepackage{color}
\usepackage{mathrsfs}
\usepackage{bbm}
\usepackage{subfigure}

\usepackage{times,txfonts}

\usepackage{siunitx}

\newcommand{\ignore}[1]{}
\newcommand{\nobibentry}[1]{{\let\nocite\ignore\bibentry{#1}}}
\newcommand{\bibfnamefont}[1]{#1}
\newcommand{\bibnamefont}[1]{#1}

\newcommand{\ket}[1]{\left\vert#1\right\rangle}
\newcommand{\bra}[1]{\left\langle#1\right\vert}
\newcommand{\meanvalue}[3]{\left\langle #1 \vert #2 \vert #3 \right\rangle}

\newcommand{\h}{\langle\hat H \rangle}
\newcommand{\hsqr}{\langle\hat H^2 \rangle}
\newcommand{\var}{\Delta\hat H^2}

\begin{document}

\title{Individual quantum probes for optimal thermometry}

\author{Luis A. Correa}
\email{LuisAlberto.Correa@uab.cat}
\affiliation{Departament de F\'{i}sica, Universitat Aut\`{o}noma de Barcelona - E08193 Bellaterra, Spain}

\author{Mohammad Mehboudi}
\affiliation{Departament de F\'{i}sica, Universitat Aut\`{o}noma de Barcelona - E08193 Bellaterra, Spain}

\author{Gerardo Adesso}
\affiliation{School of Mathematical Sciences, The University of Nottingham, University Park, Nottingham NG7 2RD, UK}

\author{Anna Sanpera}
\affiliation{Instituci\'{o} Catalana de Recerca i Estudis Avan\c{c}ats - E08011 Barcelona, Spain}
\affiliation{Departament de F\'{i}sica, Universitat Aut\`{o}noma de Barcelona - E08193 Bellaterra, Spain}

\date{\today}

\begin{abstract}
The unknown temperature of a sample can be estimated with minimal disturbance by putting it in thermal contact with an individual quantum probe. If the interaction time is sufficiently long so that the probe thermalizes, the temperature can be read out directly from its steady state. Here we prove that the optimal quantum probe, acting as a thermometer with maximal thermal sensitivity, is an effective two-level atom with a maximally degenerate excited state. When the total interaction time is insufficient to produce full thermalization, we optimize the estimation protocol by breaking it down into sequential stages of probe preparation, thermal contact and measurement. We observe that frequently interrogated probes initialized in the ground state achieve the best performance. For both fully and partly thermalized thermometers, the sensitivity grows significantly with the number of levels, though optimization over their energy spectrum remains always crucial.
\end{abstract}

\maketitle

\section{Introduction}

With the advent of quantum technologies, the study of the thermodynamics of quantum devices has attracted considerable attention \cite{e15062100,1310.0683v1}. In particular, there is a growing interest in obtaining accurate temperature readings with nanometric spatial resolution \cite{neumann2013high,kucsko2013nanometre,toyli2013fluorescence}, which would pave the way towards many ground-breaking applications in medicine, biology or material science. This motivates the development of precise quantum thermometric techniques.

Recent progress in the manipulation of \textit{individual} quantum systems has made it possible to use them as temperature probes, thus minimizing the undesired disturbance on the sample. Fluorescent thermometry may be implemented, for instance, on a single quantum dot to accurately estimate the temperature of fermionic \cite{PhysRevApplied.2.024002,PhysRevApplied.2.024001} and bosonic \cite{sabin2014impurities,BraunT} reservoirs. Similarly, the ground state of colour centres in nano-diamonds has already been used as a fluorescent thermometer \cite{neumann2013high,kucsko2013nanometre,toyli2013fluorescence}, achieving precisions down to the millikelvin scale, and a spatial resolution of few hundreds of nanometers. Thermometry applied to micro-mechanical resonators \cite{PhysRevA.84.032105,PhysRevA.86.012125,PhysRevB.88.155409}, and nuclear spins \cite{raey2014thermometer} has also been subject of investigation. Other studies have focused on more fundamental questions such as the scaling of the precision of temperature estimation with the number of quantum probes \cite{PhysRevA.82.011611}, and the potential role played by coherence and entanglement in simple thermometric tasks \cite{stace2014qubit}.

In this Letter, we investigate the fundamental limitations on temperature estimation with individual quantum probes. Two complementary scenarios are considered. In the first one, we assume that the thermometer reaches thermal equilibrium with the sample. We then determine which are the optimal probes that maximize the attainable precision in the estimation of the temperature. Alternatively, we also consider the situation in which the probe does not thermalize completely due to some constraint on the total estimation time (e.g. the sample may be unstable). In this second scenario, we analyze the dissipative time evolution of the probe in order to optimize the thermometric protocol. We model it as sequence of steps of preparation, thermal contact and readout.

Our main results are the following. First, we show that a $N$-dimensional equilibrium probe with maximum \textit{heat capacity} is optimal for thermometry. This is an effective two-level probe with $(N-1)$-degeneracy in the excited state, and some optimal gap. The maximum achievable precision grows with the dimension of the probe, yet the range of temperatures for which it operates efficiently as a thermometer becomes narrower. In contrast, a less sensitive probe with equispaced energy spectrum, such as a quantum harmonic oscillator, features wider operation ranges. On the other hand, when the estimation time is limited, we find that a frequently measured probe initialized in its  ground state achieves the largest thermal sensitivity. In this case, the overall precision still scales with the dimension of the probe, even though the temperature range for efficient operation is dimension-independent.

Our results contribute not only to the theoretical advance of temperature estimation in the quantum regime, but also have potential technological impact for the development of high precision thermometry at the nanoscale.

\section{Fully thermalized thermometers}

In standard thermometry, a (sufficiently small) thermometer is simply allowed to equilibrate with the sample to be probed, so that the temperature of the latter is inferred from the state of the probe. In a quantum scenario, the same procedure can be applied. A first approximation to the sample temperature can be obtained by performing a suitable measurement on the steady state of the thermalized probe. If a large number $\nu$ of such independent experiments is carried out, one can refine the estimate $T$ of the sample temperature. Its corresponding uncertainty $\Delta T$ is bounded from below by a geometric quantity $\mathcal{F}(\hat\varrho_T)$, known as quantum Fisher information (QFI) \cite{giovannetti2011advances},  via the quantum Cram\'{e}r-Rao inequality \cite{cramer1999mathematical,CavesRao}
\begin{equation}
\Delta T\geq[\nu \mathcal{F}(\hat\varrho_T)]^{-1/2}.
\label{cramer-rao}\end{equation}
In the present context of temperature estimation, the QFI can be interpreted as the infinitesimal distance, according to the Bures metric, between a thermal state at temperature $T$, and a thermal state at temperature $T + \delta$ \cite{CavesRao}. Intuitively, the more such a distance, the more the initial probe state is sensitive to a small variation of temperature. Formally,
\begin{equation}
\mathcal{F}(\hat\varrho_T)=-2\lim_{\delta\rightarrow 0}{\partial^2 \mathbb{F}(\hat\varrho_T,\hat\varrho_{T+\delta})}/{\partial\delta^2},
\label{fisher}\end{equation}
where $\mathbb{F}(\hat\varrho_1,\hat\varrho_2)\equiv \bigg( \mbox{tr}\,\sqrt{\sqrt{\hat\varrho_1}\hat\varrho_2\sqrt{\hat\varrho_1}}\bigg)^2$ is the Uhlmann fidelity between states $\hat\varrho_1$ and $\hat\varrho_2$, which defines their respective Bures distance via $d_{\text{Bures}}(\hat{\varrho}_1,\hat{\varrho}_2) = 2\big(1-\sqrt{\mathbb{F}(\hat\varrho_1,\hat\varrho_2)}\big)$ \cite{CavesRao}. Further to the intuitive meaning of the QFI, we note that there exists an optimal estimator (i.e., an optimal measurement procedure on the final thermalized state) for which the bound in eq.~\eqref{cramer-rao} becomes tight for an asymptotically large number of measurements ($\nu \gg 1$), and can be indeed saturated by means of adaptive metrological schemes \cite{giovannetti2011advances}. Therefore, the inverse of the QFI equivalently defines the minimum achievable variance in the estimation of $T$. We will then refer to $\mathcal{F}(\varrho_T)$ as `thermal sensitivity', and take its maximization as synonym of optimality in the following analysis \cite{PhysRevA.82.011611,PhysRevA.84.032105,PhysRevA.86.012125,sabin2014impurities,FaberT,BraunT}.

We write the Hamiltonian of our probe as $\hat H=\sum_n \epsilon_n\ket{\epsilon_n}\bra{\epsilon_n}$. A thermalization process leads to stationary states of the form $\hat\varrho_T=\sum_n p_n\ket{\epsilon_n}\bra{\epsilon_n}$, where the populations are $p_n\equiv Z^{-1}e^{-\epsilon_n/k_B T}$ and the partition function is given by $Z\equiv\mbox{tr}\,e^{-\hat H/k_B T}$. In what follows we set $\hbar=k_B=1$.

In the energy eigenbasis, eq.~\eqref{fisher} rewrites as \cite{0253-6102-61-1-08}
\begin{equation}
\mathcal{F}(\hat\varrho_T)=4{\sum}_{m,n} p_m\frac{\vert\meanvalue{\epsilon_m}{\partial_T \hat\varrho_T}{\epsilon_n}\vert^2}{(p_m+ p_n)^2}
=\frac{\Delta\hat H^2}{T^4},
\label{fisher_expl}
\end{equation}
were $\Delta \hat H^2\equiv\langle\hat H^2\rangle-\langle\hat H\rangle^2$. In this last step, we have used the identity $\langle\hat H\rangle=T^2 \partial_T \ln Z$. Interestingly, in the single shot scenario of $\nu=1$, one can combine eqs.~\eqref{cramer-rao} and \eqref{fisher_expl} to get the thermodynamic uncertainty relation $ \frac{\Delta T}{T^2}\Delta\hat H\geq 1$. Also, note that $\Delta\hat H^2/T^2=d\langle\hat H\rangle/dT\equiv C(T)$ which, in the present case, may be referred to as the `heat capacity' of the probe. It thus follows that the signal-to-noise ratio $T/\Delta T$ is upper-bounded as $(T/\Delta T)^2\leq C(T)$ \cite{PhysRevE.83.011109}. Note as well that, since $\hat\varrho_T$ is a thermal state, the most informative measurement saturating eq.~\eqref{cramer-rao} is just a projection onto the energy eigenbasis.

In the light of eq.~\eqref{fisher_expl}, the maximization of the thermal sensitivity of a probe translates into finding the energy spectrum with the largest possible energy variance at thermal equilibrium, or equivalently, the $N$-dimensional probe with largest heat capacity. Note that the heat capacity of the sample must be anyway much larger than that of the probe so as to minimize any disturbance arising from the estimation procedure.

For a general $N$-level probe, the energy variance writes as $\Delta\hat H^2 = Z^{-1} \sum_i^N \epsilon_i^2 e^{-\epsilon_i/T}-(Z^{-1} \sum_i^N \epsilon_i e^{-\epsilon_i/T})^2$, where the partition function is $Z=\sum_i e^{-\epsilon_i/T}$. The variance is bounded. In order to identify its maximum, we impose $\partial_{\epsilon_i}\Delta H^2=0$, which results in a set of $N$ transcendental equations. Subtracting the $j$-th equation from the $i$-th one ($\partial_{\epsilon_i}\Delta\hat H^2-\partial_{\epsilon_j}\Delta\hat H^2=0$), we arrive at the condition $(\epsilon_i-\epsilon_j)(\epsilon_i+\epsilon_j-2-2\langle\hat H\rangle/T)=0$ (see \cite{suppl} for details). That is, any two energy eigenvalues $\epsilon_i$ and $\epsilon_j$ must be either equal, or sum up to the same value at the stationary points of $\Delta\hat H^2$. This may only happen if the energy spectrum is that of an effective two-level atom with energies $\{\epsilon_-,\epsilon_+\}$, and $N_0$ and $N-N_0$ times degenerate ground and excited state, respectively. Without loss of generality, we may always shift the energy spectrum so that $\epsilon_-=0$ and the optimal gap becomes $x^*_{N,N_0}\equiv\Omega^*/T= (\epsilon_+ + \epsilon_-)/T=2(1+\langle\hat H\rangle/T)>2$, since now $\langle\hat H\rangle>0$. This critical gap may be conveniently rewritten as $e^{x^*_{N,N_0}}=\frac{N-N_0}{N_0}\frac{x^*_{N,N_0}+2}{x^*_{N,N_0}-2}$. Observing that the difference $\Delta \hat H^2(x^*_{N,N_0-1};N,N_0-1)-\Delta \hat H^2(x^*_{N,N_0};N,N_0)=\frac14({x^{*2}_{N,N_0-1}}-{x^{*2}_{N,N_0}})$ is always positive, one can conclude that the excited-state degeneracy must be the largest possible (i.e.~$N_0=1$) so as to maximize the energy variance. 

Finally, to ensure that $\Delta\hat H^2$ reaches a maximum at $x^*_{N,N_0}$, we must check that the Hessian matrix ($\mathcal{H}_{ij}\equiv\partial^2\Delta\hat H^2/\partial\epsilon_i\partial\epsilon_j$) is negative definite in that configuration. After a tedious but otherwise straightforward calculation, we can see that it has $N-2$ identical eigenvalues $\lambda_1=-\frac{1}{2}\frac{x^*_{N,1}-2}{N-1}$, plus two non-degenerate ones: $\lambda_2=-\frac18 \frac{x_{N,1}^{*2}-4}{N-1}$ and $\lambda_3=0$. Since $x_{N,1}^*>2$, both $\lambda_1$ and $\lambda_2$ are negative. The single vanishing eigenvalue $\lambda_3$ simply reflects the obvious symmetry of $\Delta\hat H^2$ with respect to a global shift of all energy levels. Hence, one may rigorously conclude that the effective two-level configuration described above indeed maximizes the energy variance. Note that this is in agreement with \cite{1304.0036}.

\begin{figure}
	\includegraphics[width=0.8\columnwidth]{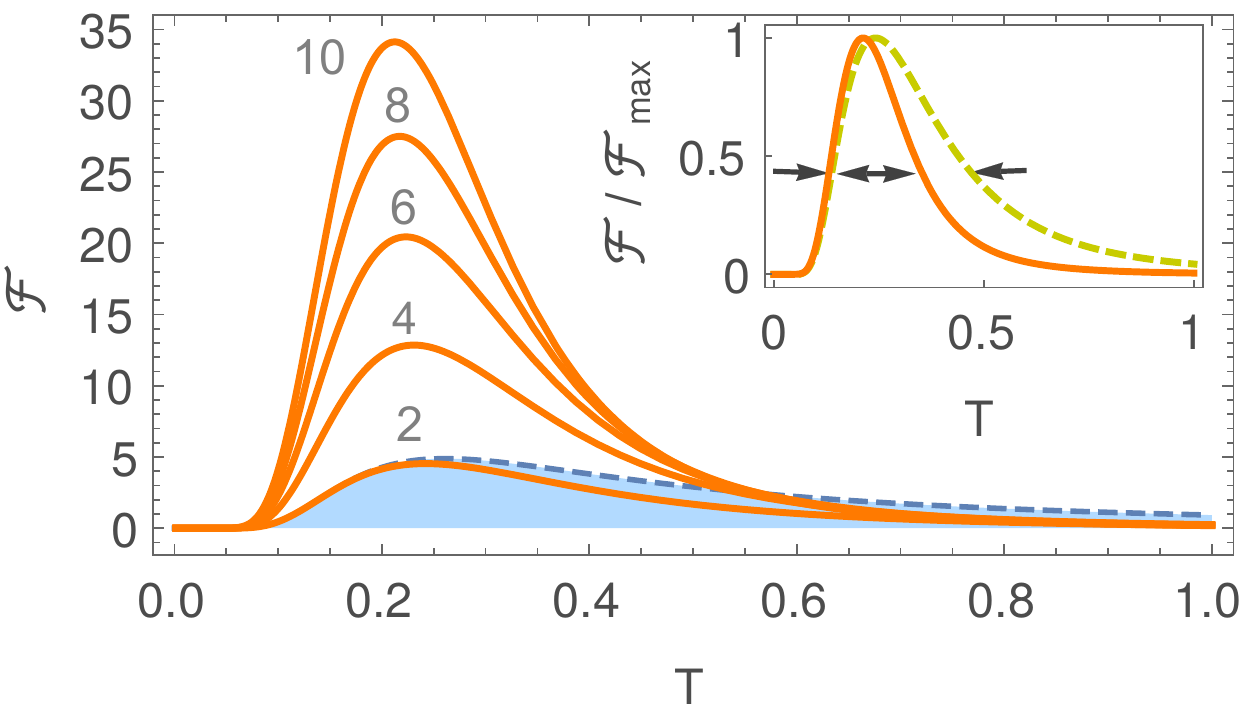}
	\caption{QFI versus sample temperature for optimized $N$-dimensional probes (orange) with $N=\{2,4,6,8,10\}$. The dashed blue line represents the QFI of a harmonic probe and the shaded blue area is the domain reachable by finite-dimensional probes with equispaced spectrum. In the inset, the normalized sensitivities of two probes with $N=2$ (dashed green) and $N=10$ (solid orange) are compared. The arrows indicate the width of the specified temperature range. Temperature and QFI are both expressed in arbitrary units and $\Omega=1$.}
\label{fig1}
\end{figure}

Here is the final expression for the corresponding QFI
\begin{equation}
\mathcal{F}_N=\frac{x ^4 \, e^{x}}{\Omega^2}~\frac{N-1}{\left(N-1+e^{x}\right)^2},
\label{fisher_opt}\end{equation}
which is obviously also maximal at $x=x^*_{N,1}$. In fig.~\ref{fig1}, we plot eq.~\eqref{fisher_opt} for different values of $N$. The precision in temperature estimation improves significantly by increasing the dimensionality $N$ of the probe, albeit at the expense of reducing the specified temperature range for efficient operation of the probe as a thermometer (see inset of fig.~\ref{fig1}).

So far, we have seen that the best thermometers are effective two-level atoms with a highly degenerate excited state and a specific, temperature-dependent gap. However, these may be very hard to prepare in practice, especially due to the fact that the sample temperature must be known precisely. For this reason we now consider more versatile sub-optimal probes with a richer spectrum, such as a single thermalized harmonic oscillator. In this case, the corresponding QFI can be easily computed from the $2\times 2$ steady-state \textit{covariance matrix} \cite{0503237v1,adesso2014continuous} of a thermal state $\boldsymbol\sigma_T=\coth{\frac{\Omega}{2T}}\mathbbm{1}_{2}$ as in eq.~\eqref{fisher}. Using the fact that the Uhlmann fidelity between two single-mode Gaussian states $\boldsymbol\sigma_1$ and $\boldsymbol\sigma_2$ is given by $\mathbb{F}(\boldsymbol\sigma_1,\boldsymbol\sigma_2)= 2\big(\sqrt{\Delta + \Lambda}-\sqrt{\Lambda}\big)^{-1}$ \cite{0305-4470-31-15-025}, where $\Delta\equiv\text{det}(\boldsymbol{\sigma}_1+\boldsymbol{\sigma}_2)$ and $\Lambda\equiv\text{det}(\boldsymbol{\sigma}_1-1)\text{det}(\boldsymbol{\sigma}_2-1)$, one arrives at $\mathcal{F}_{\text{ho}}=\frac{\Omega ^2}{4 T^4}~\text{csch}^2\frac{\Omega }{2 T}$. This is represented in fig.~\ref{fig1} with a dashed blue line. For ease of comparison we take the oscillator frequency $\Omega$ to be $\epsilon_+-\epsilon_-$. As we can see, a harmonic probe features a thermal sensitivity similar to that of a two-level probe. Even if harmonic thermometers are  outperformed by most optimized $N$-level probes, they are endowed with a much broader specified temperature range for efficient operation, making them a choice of practical interest. This can be understood by observing that the thermal sensitivity of a probe with a single energy gap may only peak at one characteristic frequency, while with an equispaced, unbounded spectrum there will always be some transition close to resonance.

\section{Partly thermalized thermometers}

All the previous analysis holds regardless of the probe-sample interactions or the spectral properties of the sample, as long as thermalization takes place. In practice, however, one may have to read out the temperature \textit{before} attaining full thermalization. This would be the case, for instance, if the sample was unstable and existed only for times comparable to the dissipation time scale.
In this alternative scenario, we ask ourselves about the optimal breakup of the total running time of the estimation procedure ($t_s$) into sequential stages of probe-preparation, thermal contact (during time $\Delta t$), and measurement, so as to optimize the achievable precision in eq.~\eqref{cramer-rao}. Note that the number of interrogations is now limited to $\nu=t_s/\Delta t$, so that the figure of merit to be maximized is the ratio $\mathcal{F}(\Delta t)/\Delta t$ \cite{PhysRevLett.79.3865,PhysRevLett.109.233601}.

Since we must monitor the time evolution of the probe, it is necessary now to specify the sample and its coupling with the thermometer. We shall model the sample as a bosonic heat bath, linearly coupled to an arbitrary probe. The total Hamiltonian writes as $\hat H_{\text{tot}}=\hat H+\sum_\mu \omega_\mu~\hat b^\dagger_\mu \hat b_\mu+\hat{X}\otimes\sum_\mu g_\mu (\hat b_\mu+\hat b_\mu^\dagger)$, where $\hat b_\mu$ is the annihilation operator of mode $\omega_\mu$ in the sample. We choose the probe-sample coupling constants to be $g_\mu=(\gamma \omega_\mu)^{1/2}$, implying flat spectral density $J(\omega)\sim \sum_\mu \frac{g_\mu^2}{\omega_\mu}\,\delta(\omega-\omega_\mu)=\gamma$ \cite{weiss2008quantum}. This sets the time-scale $\tau_D\sim\gamma^{-1}$ over which $\hat\varrho(t)$ varies appreciably.
Tracing out the sample from the overall unitary dynamics and assuming a thermal state $\hat \chi_T$ for it, leads to an effective equation of motion of the Lindblad-Gorini-Kossakovski-Sudashan type (LGKS) \cite{lindblad1976generators,gorini1976completely}, that follows from $\dot{\hat\varrho}=\mbox{tr}_S\frac{d}{dt}\{e^{-i\hat H_{\text{tot}}t}~\hat \varrho(0)\otimes\hat \chi_T~e^{i \hat H_{\text{tot}}t}\}$, after sequentially performing the Born, Markov and rotating-wave approximations (see \cite{breuer2002theory} for a detailed derivation). Note that the Born approximation implies that no correlations are ever created between probe and sample, so the latter  remains undisturbed throughout the estimation procedure. Note also that, for consistency with the Markov approximation, the temperature of the sample may be not arbitrarily low, as the thermal fluctuations must remain fast compared with $\tau_D$.

In the interaction picture, the master equation can be cast as
\begin{multline}
\dot{\hat \varrho}=\Gamma_{\Omega,T} \left(\hat A_\Omega\hat \varrho \hat A_{-\Omega}-\mbox{$\frac{1}{2}$}\{\hat A_{-\Omega} \hat A_\Omega,\hat \varrho\}_+\right) \\
+ e^{-\Omega/T}\Gamma_{\Omega,T} \left(\hat A_{-\Omega}\hat \varrho \hat A_\Omega-\mbox{$\frac{1}{2}$}\{\hat A_\Omega \hat A_{-\Omega},\hat \varrho\}_+\right),
\label{master_equation}\end{multline}
where $\hat A_{\pm\Omega}$ stands for the relaxation/excitation operator associated with the decay channel at frequency $\Omega$. These follow from the decomposition of $\hat{X}=\sum_{\omega} \hat A_\omega$ as sum of eigen-operators of the probe Hamiltonian (i.e. such that $[\hat H, \hat A_\Omega]=-\Omega\,\hat A_\Omega$). It is easy to show that the thermal state $\hat\varrho = Z^{-1} e^{-\hat H/T}$ is a fixed point of eq.~\eqref{master_equation} and, choosing a suitable coupling operator $\hat X$, the open dynamics may also be \textit{ergodic}, thus eventually bringing any initial state to thermal equilibrium \cite{breuer2002theory}.

For a two-level thermometer with Hamiltonian $\hat H=\frac{\Omega}{2} \hat{\sigma}_z$, we can take, for instance, $\hat{X}=\hat{\sigma}_x$ from which $\hat A_\Omega=\ket{-\Omega/2}\bra{\Omega/2}$, while $\hat A_{-\Omega}=\hat A_\Omega^\dagger$. Here, $\ket{\pm\Omega/2}$ are the corresponding energy eigenstates. Generalizing to the case of an $N$-level probe with eigenstates $\{\ket{\epsilon_i}\}$, a coupling term like $\hat X=\sum_{i\neq 1}\ket{\epsilon_1}\bra{\epsilon_i}+\ket{\epsilon_i}\bra{\epsilon_1}$ would also thermalize any preparation, where we have labelled the ground state by $\ket{\epsilon_1}$. The resulting relaxation operators are $\hat A_{\epsilon_i-\epsilon_1}=\ket{\epsilon_1}\bra{\epsilon_i}$. In particular, to account for our effective two-level systems with excited-state degeneracy we can take the limit $\epsilon_i\rightarrow\frac{\Omega}{2}$ for $i\neq 1$ and set $\epsilon_1=-\frac{\Omega}{2}$ to get the desired thermalization process.
Let us finally comment on the decay rates $\Gamma_{\Omega,T}$, which follow from the power spectrum of the bath auto-correlation function $\langle \hat S(t)\hat S(0)\rangle_T\equiv\mbox{tr}\,\{\hat{S}(t)\hat{S}(0)\hat\chi_T\}$, where $\hat S\equiv\sum_\mu g_\mu (\hat b_\mu+\hat b_\mu^\dagger)$. In the specific case of a quantum probe coupled through dipole interaction to the quantized electromagnetic field in three dimensions, one obtains $\Gamma_{\Omega,T}=\gamma\Omega^3(1-e^{-\Omega/T})^{-1}$ \cite{breuer2002theory}.

The problem now goes down to  solving eq.~\eqref{master_equation}, transforming the time-evolved state $\hat \varrho(t)$ back into the Schr\"{o}dinger picture (i.e. $\hat \varrho\mapsto e^{-i \hat H t}\hat \varrho e^{i \hat H t}$), and computing the QFI according to eq.~\eqref{fisher} \cite{PhysRevLett.79.3865,escher2011noisy,demkowicz2012elusive,PhysRevLett.112.120405,1403.8033v2}. Note that besides comparing the performance of different types of probe, we must now optimize over their initial state too.
We start by considering the simplest case of two-level thermometers. Extensive numerical analysis over different initial states shows that ground-state preparations display maximal thermal sensitivity. This indicates that the presence of initial quantum coherence does not provide any significant advantage for thermometry in this setting.

\begin{figure}
\includegraphics[width=0.8\columnwidth]{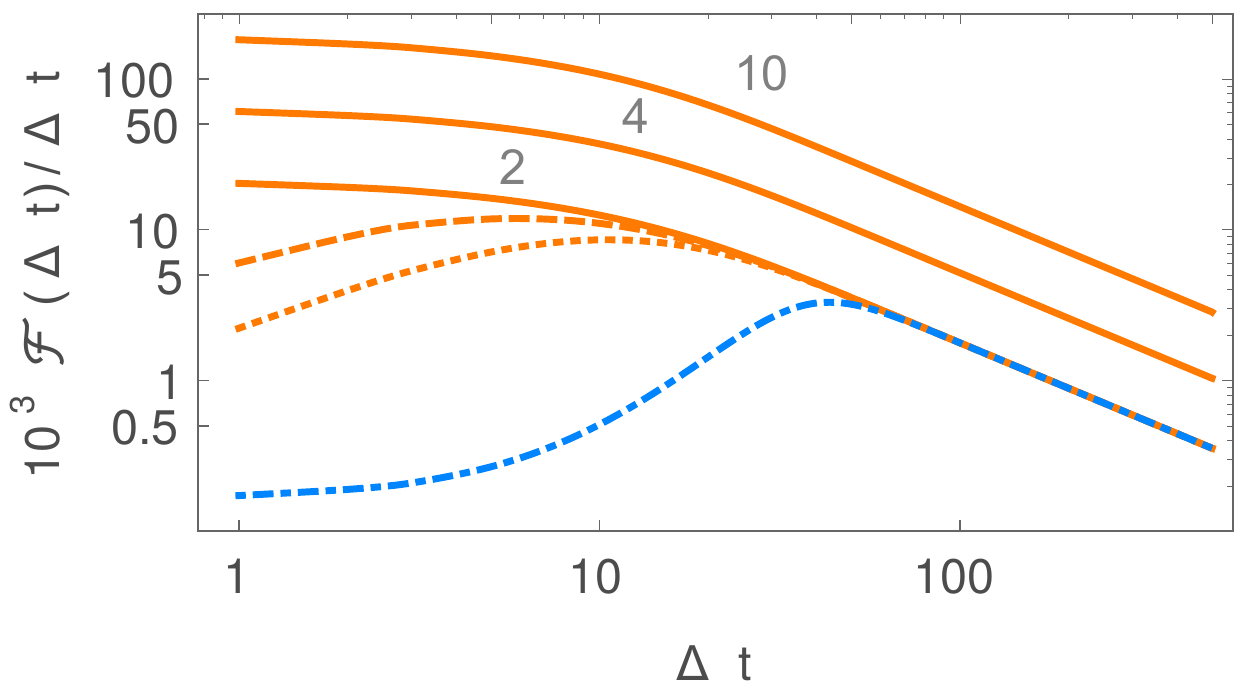}
\caption{Log-log plot of $\mathcal{F}/\Delta t$ as a function of $\Delta t$ for different preparations and probe dimensionalities. The continuous orange lines stand for $\mathcal{F}_N$ for probes with $N=\{2,4,10\}$ initialized in the ground state. The dashed and dotted orange curves stand for a two-level probe initialized in a thermal state at temperature $0.8$ and $0.9$, respectively. The dot-dashed blue curve corresponds to a two-level probe prepared in the maximally coherent state $\hat\varrho(0)=\ket{+}\bra{+}$ ($\Omega/T=\tilde{x}$, $\gamma=10^{-3}$, and $T=1$, in arbitrary units).}\label{fig2}
\end{figure}

Thus, by choosing $\hat\varrho(0)=\ket{-\Omega/2}\bra{-\Omega/2}$ we can combine eqs.~\eqref{master_equation} and \eqref{fisher_expl} to compute $\mathcal{F}_2(\Delta t)$ as a function of the interrogation time $\Delta t$, starting from a ground state preparation:
\begin{equation}
\mathcal{F}_2(\Delta t)\\=
\frac{x^2\left(e^{x}\left(e^{\Delta t/\tau}-1\right)+\left(1+e^{x}\right)\frac{\Delta t}{2\tau}\mbox{csch}\,\frac{x}{2}\right)^2}{\left(1+e^{x}\right)^2\left(e^{\Delta t/\tau}-1\right)\left(1+e^{x}e^{\Delta t/\tau}\right)T^2},
\label{TimeEvo_qbt_formula}\end{equation}
where $\tau^{-1}\equiv\gamma\,\Omega^3\coth{\frac{x}{2}}$. Eq.~\eqref{TimeEvo_qbt_formula}  shows that the details of the thermal fluctuations of the sample, encoded in $\Gamma_{\Omega,T}$, only enter in the dynamics through the scaling factor $\tau$. Hence, even if our choice of a flat spectral density might seem pretty restrictive at first, changing the probe-sample coupling would just amount to a suitable rescaling of time.

In fig.~\ref{fig2} we plot $\mathcal{F}_2(\Delta t)/\Delta t$ for different preparations. As we can see, the sensitivity of a cold thermal probe peaks at some optimal readout time, after which it must be quickly cooled down to start over another relaxation stage in the estimation protocol. In the limiting case of a ground-state preparation, the overall maximum sensitivity is approached as $\Delta t\rightarrow 0$.

Eq.~\eqref{TimeEvo_qbt_formula} can be generalized to any of our highly degenerate effective two-level probes prepared in the ground state. As before, their maximum  precision follows from  the limit
\begin{equation}
\lim_{\Delta t\rightarrow 0}\frac{\mathcal{F}_N(\Delta t)}{\Delta t}=\frac{\gamma  T (N-1) x^5 e^{2x}}{\left(e^{x}-1\right)^3}.
\label{ultimate_limit}\end{equation}
We now search for the optimal frequency-to-temperature ratio $\tilde{x}$ that sets an ultimate upper bound on the thermal sensitivity in eq.~\eqref{ultimate_limit}. This can be expressed implicitly as $e^{\tilde{x}}=(5+2\tilde{x})/(5-\tilde{x})$, which is independent of $N$. Interestingly, the specified temperature range for efficient operation does not scale with $N$, at variance with the fully thermalized case.

For completeness, we examine again here the performance of harmonic probes. Going back to eq.~\eqref{master_equation}, we will set $\hat{H}=\Omega \, \hat{a}^\dagger \hat{a}$ and $\hat{X}=\hat{a}+\hat{a}^\dagger$, whose corresponding relaxation and excitation operators are trivially $\hat A_\Omega= \hat a$ and $\hat A_{-\Omega}= \hat a^\dagger$. The total Hamiltonian is thus quadratic in positions and momenta and therefore, any Gaussian preparation will preserve its Gaussianity in time \cite{0503237v1}. Provided that the initial state also has vanishing first order moments ($\left\langle\hat x\right\rangle=\left\langle\hat p\right\rangle=0$), its covariance matrix $\boldsymbol{\sigma}(t)$ alone will be enough for a full description.

In this case, the dynamics may be obtained by explicitly solving the quantum master equation in phase space, to yield $\boldsymbol\sigma(t)=e^{-\Gamma_{\Omega,T}t}\boldsymbol\sigma(0)+(1-e^{-\Gamma_{\Omega,T}t})\boldsymbol\sigma_T$ \cite{PhysRevA.68.012314,0503237v1}. Computing the transient QFI is thus straightforward by resorting to eq.~\eqref{fisher}. In what follows, we shall consider general (undisplaced) single-mode Gaussian states as initial preparations; these can be written as rotated, squeezed thermal states \cite{0503237v1,adesso2014continuous}. As it could be expected, ground-state initialization ($\hat\varrho(0)=\ket{0}\bra{0}$) provides once again the largest thermal sensitivity.
One can ignore the temperature dependence of $\Gamma_{\Omega,T}$ in the solution to the master equation and still get a good approximation to $\lim_{\Delta t\rightarrow 0}\mathcal{F}_\text{ho}(\Delta t)/\Delta t$. Surprisingly, we recover eq.~\eqref{ultimate_limit} with $N=2$. Indeed, this equivalence of two-level probes and harmonic thermometers extends generally beyond the limits $\Delta t\rightarrow 0$ and $\hat\varrho(0)=\ket{0}\bra{0}$. Therefore, at variance with the fully-thermalized scenario, the specified temperature range of both oscillators and $N$-level probes in an effective two-level configuration is virtually the same, regardless of $N$.

\section{Conclusions}

We have analyzed the performance and ultimate limitations of individual quantum probes for precise thermometry on a sample. Our study is based on techniques of  parameter estimation \cite{giovannetti2011advances,CavesRao}, and makes use of the quantum Fisher information as indicator of optimal thermal sensitivity.

First, we have considered a general $N$-dimensional quantum probe that fully thermalizes with the sample. We have  linked the quantum Fisher information with the heat capacity of the probe, and proven that the best quantum thermometer is an effective two-level atom with a maximally degenerate excited state at a specific energy gap, depending non-trivially on the sample temperature. There exists a complementary trade-off  between the maximum achievable estimation precision, which grows with  $N$, and the specified temperature range in which the estimation is efficient, which shrinks with  $N$.

We have also considered the scenario in which, e.g.~due to short lifetime of the sample, full thermalization may not take place.
Frequently interrogated probes prepared in their ground state then provide the largest thermal sensitivity. While the maximum achievable precision scales again with $N$, the specified temperature range is dimension-independent in this case. These results were obtained by considering a large bosonic sample in thermal equilibrium, weakly coupled to the probe through a linear interaction term, ensuring ergodicity. It would
be interesting to discuss to which extent can the estimation precision be enhanced with a suitably engineered thermal coupling, e.g. by externally controlling 
the scattering length in a cold atomic gas \cite{1105.4790v1}. In principle, this would allow the experimenter to directly manipulate the scaling factor $\tau$ in eq.~\eqref{TimeEvo_qbt_formula}.

Finally, it is worth mentioning that even though quantum coherence in the initial state of the probe may not be directly linked to the overall maximization of the precision, the potential role played by \textit{quantumness} in thermometry remains an open problem \cite{stace2014qubit,raey2014thermometer} that deserves a study on its own.

\acknowledgments

The authors would like to thank J. Calsamiglia, J. Filgueiras, T. Bromley and M. Cianciaruso for fruitful discussions, and K. Hovhannisyan for making us aware of ref.~\cite{1304.0036} after completion of this work. Financial support from Spanish MINECO (FIS2008-01236), EU Collaborative Project TherMiQ (Grant Agreement No. 618074), European Regional Development Fund, COST Action MP1209, Generalitat de Catalunya (Grant No. SGR2014-966), Brazilian CAPES (Grant No. 108/2012), Foundational Questions Institute (Grant No. FQXi-RFP3-1317), and ERC StG GQCOP (Grant Agreement No. 637352) is  acknowledged.

\appendix

\section{Details on the proof of the optimality of the effective two-level probes}

Below, we give further details on the proof of the optimality of effective two-level thermalized probes with $N-1$ times degenerate excited state, for the maximization of the energy variance. For an $N$-level probe in a thermal state, this writes as
\begin{widetext}
\begin{equation}
\Delta\hat H^2(\{\epsilon_i\})=\langle\hat H^2\rangle-\langle\hat H\rangle^2=Z^{-1}\sum_{i=1}^N\epsilon_i^2 e^{-\epsilon_i/T}-\left(Z^{-1}\sum_{i=1}^N\epsilon_i e^{-\epsilon_i/T}\right)^2,
\label{s_variance}\end{equation}
where $Z=\sum_i e^{-\epsilon_i/T}$. In order to find the stationary points of $\Delta\hat H^2(\{\epsilon_i\})$ we simultaneously impose the $N$ conditions $\partial_{\epsilon_i}\Delta\hat H^2(\{\varepsilon_i\})=0$, which result in the following system of transcendental equations 
\end{widetext}
\begin{widetext}
\begin{equation}
\frac{e^{-\epsilon_i/T}}{Z}\left[\frac1T\left(\langle\hat H^2\rangle-2\langle\hat H\rangle^2\right)+\epsilon_i\left(2-\frac{\epsilon_i}{T}\right)+2\langle\hat H\rangle\left(\frac{\epsilon_i}{T}-1\right)\right]=0\qquad\forall i\in\{1,\cdots,N\}.
\label{s_first_der}\end{equation}
\end{widetext}
One may now subtract the $j$-th of such equations from the $i$-th one, obtaining
\begin{equation}
(\epsilon_i-\epsilon_j)\left[\epsilon_i+\epsilon_j-2\left(\langle\hat H\rangle+T\right)\right]=0.
\label{s_difference}\end{equation}
That is, the stationary points of $\Delta\hat H^2$ are such that any two energy eigenvalues must be either equal or sum up to $2\big(\langle\hat H\rangle+T\big)$. A set of conditions like eq.~\eqref{s_difference} cannot be simultaneously met by more than two different energy eigenvalues $\{\epsilon_+,\epsilon_-\}$. Hence, the only energy spectra compatible with stationarity are those of effective two-level atoms with ground state degeneracy $N_01$ and an $N-N_0$ times degenerate excited state. Without loss of generality, we may always shift the spectrum so as to set $\epsilon_-=0$. According to eq.~\eqref{s_difference}, the gap of the effective two-level system becomes $\Omega^*\equiv\epsilon_+-\epsilon_-=2\big(\langle\hat H\rangle+T\big)$. 

Note that for an effective two-level probe the average energy rewrites as
\begin{equation}
\langle\hat H(x;N,N_0)\rangle=T\frac{(N-N_0)x\,e^{-x}}{N_0+(N-N_0)\,e^{-x}},
\end{equation}
where we have introduced the notation $x\equiv\Omega/T$ for the frequency-to-temperature ratio (recall that we work in units of $\hbar=k_B=1$). The energy gap at stationarity can be thus conveniently cast as 
\begin{equation}
e^{x^*_{N,N_0}}=\frac{N-N_0}{N_0}\frac{x^*_{N,N_0}+2}{x^*_{N,N_0}-2}.
\label{s_optimal_gap}\end{equation}
In order to determine the ground and excited state degeneracies yielding the largest energy variance at the critical frequency-to-temperature ratio, we can compare $\Delta\hat H^2(x_{N,N_0}^*;N,N_0)$ with $\Delta\hat H^2(x_{N_0-1,N}^*;N,N_0-1)$, where
\begin{equation}
\Delta\hat H^2(x;N,N_0)=T^2\frac{N_0(N-N_0)\,x^2e^x}{\left[(N-N_0)+N_0\,e^x\right]^2}.
\end{equation} 
This yields $\Delta \hat H^2(x^*_{N,N_0-1};N,N_0-1)-\Delta \hat H^2(x^*_{N,N_0};N,N_0)=\frac14({x^{*2}_{N,N_0-1}}-{x^{*2}_{N,N_0}})>0$, which is positive according to eq.~\ref{s_optimal_gap}. Hence, an effective two-level probe with maximally degenerate excited state (i.e. $N_0=1$) has the largest energy variance at stationarity.

All that is left is to prove that such stationary point is indeed a maximum for the energy variance. For that purpose, we shall compute explicitly the elements of the Hessian matrix $\mathcal{H}_{ij}\equiv\partial^2\Delta\hat H^2/\partial\epsilon_i\partial\epsilon_j$ and check its eigenvalues for negative definiteness at the stationary point.
After a lengthy but otherwise straightforward calculation, one can see that the diagonal elements evaluate to

\begin{widetext}
\begin{multline}
\mathcal{H}_{ii}=\frac{\partial^2\Delta\hat H^2}{\partial\epsilon_i^2}=\left(\frac{e^{-\epsilon_i/T}}{Z\,T}\right)^2\left[2\big(\hsqr-3\h^2-4T\h-T^2\big)+8(T+\h)\epsilon_i-4\epsilon_i^2 \right. \\
\left.+Z\,e^{\epsilon_i/T}\big(2T^2+2\h\big(2T+\h\big)-\hsqr-2\big(2T+\h\big)\epsilon_i+\epsilon_i^2\big)\right]\qquad\forall i\in\{2,\cdots,N\},
\label{s_diagonal}\end{multline}
\end{widetext}
while the off-diagonals are given by
\begin{widetext}
\begin{equation}
\mathcal{H}_{ij}=\frac{\partial^2\var}{\partial\epsilon_i\partial\epsilon_j}=
\frac{e^{-(\epsilon_i+\epsilon_j)/T}}{T^2 Z^2}\left[4\h(\epsilon_i+\epsilon_j-2T)+(\epsilon_i+\epsilon_j)(4T-\epsilon_i-\epsilon_j)+2\hsqr-6\h^2-2T^2\right]\qquad\forall i\neq j.
\label{s_offdiagonal}\end{equation}
\end{widetext} 
We are interested in the particular case of an effective two-level spectrum with $N-1$ times degenerate excited state at the corresponding optimal gap $x^*\equiv x_{N,1}^*$. For this configuration, the Hessian has the following structure
\begin{equation}
\boldsymbol{\mathcal{H}} =
 \begin{pmatrix}
  a & c & c & \cdots & c \\
  c & b & d & \cdots & d \\
  c & d & b & \cdots & d \\
  \vdots  & \vdots  & \vdots & \ddots & \vdots  \\
  c & d & d & \cdots & b
 \end{pmatrix},
\label{s_hessian}\end{equation}
where $a=\mathcal{H}_{ii}\vert_{\epsilon_i=0}$, $b=\mathcal{H}_{ii}\vert_{\epsilon_i=x^*}$, $c\equiv\mathcal{H}_{ij}\vert_{\epsilon_i=0,\epsilon_j=x^*}=\mathcal{H}_{ij}\vert_{\epsilon_i=x^*,\epsilon_j=0}$ and $d=\mathcal{H}_{ij}\vert_{\epsilon_i=\epsilon_j=x^*}$. To compute these elements from eqs.~\eqref{s_diagonal} and \eqref{s_offdiagonal} we can also make the following replacements $Z=2x^*/(2+x^*)$, $\h=\frac{T}{2}(x^*-2)$ and  $\hsqr=\frac{T^2}{2}x^*(x^*-2)$ and $e^{x^*}=(N-1)\frac{x^*+2}{x^*-2}$., yielding
\begin{subequations}
\begin{align}
a &= -\frac{1}{8}(x^{*2}-4) \qquad \\
b &= -\frac{(x^*-2)(4N-6+x^*)}{8(N-1)^2} \qquad \\
c &=\frac{(x^{*2}-4)^2}{8(N-1)} \qquad \\
d &=-\frac{(x^*-2)^2}{8(N-1)^2} \,. 
\label{s_simplified}\end{align}
\end{subequations}
Finally, the diagonalization of the Hessian leads to the following eigenvalues: $\lambda_1=-(x^*-2)/(2(N-1))$ ($N-2$ times degenerate), $\lambda_2=-(x^{*2}-4)/(8(N-1))$ and $\lambda_3=0$ (both non-degenerate). The vanishing eigenvalue follows from the invariance of $\Delta\hat H^2$ under uniform global shifts of all energy levels. Note as well that $x^*>2$, as follows from eq.~\eqref{s_optimal_gap}, implying that both $\lambda_1$ and $\lambda_2$ are negative definite. This demonstrates that the stationary point corresponds indeed to a maximum. 

We have thus rigorously proven that an effective two-level spectrum with an $N-1$ times degenerate excited state at energy $T\,x^*$ yields the largest possible energy variance for an $N$-dimensional system at thermal equilibrium with temperature $T$. $\blacksquare$

\bibliographystyle{apsrev}

\end{document}